\documentstyle[aps,twocolumn,epsfig]{revtex}

\begin{document}

\newcommand{\ve}[1]{\mbox{\boldmath $#1$}}
\twocolumn[\hsize\textwidth\columnwidth\hsize
\csname@twocolumnfalse%
\endcsname
\draft
\title{Vortices in superfluid trapped Fermi gases at zero temperature}

\author{G. M. Bruun$^1$ and L. Viverit$^2$}
\address{$^1$ Nordita, Blegdamsvej 17, 2100 Copenhagen, Denmark}
\address{$^2$ Dipartimento di Fisica, Universit\`a di Milano, via Celoria 16,
20133 Milano, Italia}

\maketitle

\begin{abstract}
\noindent 
We discuss various aspects of the vortex state of a dilute 
superfluid atomic Fermi gas at $T=0$. 
The energy of the vortex in a trapped gas is calculated 
and we provide an expression for the 
thermodynamic critical rotation frequency of the trap 
for its formation. Furthermore, 
we propose a method to detect the presence of a vortex 
by calculating the effect of its associated  velocity field 
on the collective mode spectrum of the gas.
\end{abstract}

\vskip2pc]

Inspired by the impressive progress in recent years in the 
field of Bose-Einstein condensation in dilute atomic gases,
increasing attention  is being devoted to examine the behavior 
of a  gas of fermionic atoms at the same 
ultra-low temperatures. 
Experimentally, the  trapping and cooling of fermionic 
alkalis has been demonstrated reaching temperatures 
as low as $\sim T_F/4$   for  $^{40}$K~\cite{DeMarco} 
and $^6$Li~\cite{Truscott,Schreck,O'Hara} with 
$T_F$ denoting the Fermi temperature. It is well known from 
condensed matter and nuclear physics that 
a gas composed of 
two different internal states of the same fermionic particle 
which interact via an attractive interaction is
unstable to formation of so-called Cooper pairs, thus becoming a superfluid. 
After the possibility of such superfluid transition for 
trapped Fermi gases was proposed~\cite{Stoof}, a lot 
of theoretical work has been focusing on various properties 
of this system~\cite{Fermipapers}. 
At the same time a major experimental goal has become to 
observe the formation of the superfluid state.

One of the intriguing properties of a superfluid is the 
possibility of forming quantized vortices. For a 
Bose--Einstein condensate, the study of vortices has 
produced several interesting results~\cite{Fetterreview}.
Recently, some aspects of the  vortex state of a trapped 
superfluid Fermi gas close to the critical temperature $T_c$ of 
the superfluid phase transition were considered~\cite{Rodriguez}.
In this paper we are interested in the properties of the vortex 
state of clouds of trapped Cooper-paired 
fermions at $T=0$, and 
in particular in understanding under which conditions a vortex 
forms, what is its energy,  and how it can be detected.
 We consider large systems where 
the coherence length $\xi$ of the superfluid is much smaller than the extent
 of the cloud. In this limit, we are 
able to derive an analytical estimate of  the energy of a vortex in a trap,
thereby predicting the  critical rotation frequency for its formation.
 Also, we propose a way of  observing the vortex by 
calculating its effect on the collective mode 
spectrum of the gas.   The paper is organized as follows: 
first in Sec.\ \ref{Coherencelength}, we examine for 
which values of the characteristic parameters the vortex is 
well localized within the gas.
In Sec.\ \ref{Uniformvortex} we present a simple model 
for calculating the energy of a vortex in a uniform superfluid Fermi system.
Using the result of Sec.\ \ref{Uniformvortex}, 
in Sec.\ \ref{Trappedvortex} we calculate the 
energy of a singly quantized vortex in a trapped gas and from that
we obtain the value of the thermodynamic critical rotation frequency
for its formation.
The problem of observing the vortex state is
considered in Sec.\ \ref{Observation}, where we calculate 
how the presence of a vortex influences the collective
mode spectrum of the gas. Finally, we summarize our results 
in Sec.\ \ref{Conclusion}.
Given the uncertainties intrinsic in any simple model of the vortex,
such as the one presented in Sec.\ \ref{Uniformvortex},
in App.\ \ref{Ginzburg} we briefly discuss another possible way of 
describing a vortex in a uniform gas, based on an 
approximate zero--temperature
Ginzburg--Landau approach \cite{epstbaym}. We then calculate what
this alternative method gives for the energy of the vortex, and compare
the two results to show that they do not differ in any significant way.

\section{Basic considerations}\label{Coherencelength}

In the dilute ultracold limit
the effective interaction between identical fermionic 
atoms vanishes due to the Pauli principle, and that between
different ones can 
be well described by one parameter only, the $s$-wave scattering length $a$. 
For a negative scattering length, 
the interaction is attractive and if
the number of particles in the two internal states is the same
the $T=0$ ground state of the gas is a superfluid. 
The critical temperature $T_c$
for the transition to such a superfluid state in a dilute gas 
was first determined for a uniform system by Gorkov
and Melik-Barkhudarov \cite{gorkov}, and using a more modern approach by 
Heiselberg {\em et al.} \cite{heiselberg}. The predicted value is 
\begin{equation}
k_BT_c=\frac{\gamma}{\pi}\left(\frac{2}{e}\right)^{7/3}\epsilon_F\;
e^{-1/\lambda},
\label{tbcs}
\end{equation} 
where $\lambda$ stands for $2k_F |a|/\pi$, $\epsilon_F$ is the Fermi 
energy common to the two species of 
fermions, $k_F$ the associated Fermi wavenumber, 
and $\gamma\simeq 1.781$ is related to Euler's constant $C$ by $\gamma=e^C$. 
The pairing gap $\Delta$ at $T=0$ is, as usual in BCS theory, related
to the critical temperature by
$\Delta_0=\pi\gamma^{-1}k_BT_c$~\cite{degennes,fetter}. 

When applying this result to a gas trapped by a harmonic oscillator 
potential, as in the cases of experimental interest today, 
some requirements have to be met. 
The first one is, just as for the uniform case, 
that the density is everywhere so low that the gas is dilute, i.e.\ $k_F({\bf r})|a|\ll 1$. We have
introduced a local Fermi wavenumber $k_F({\bf r})$. This corresponds to using
the Thomas-Fermi approximation, which is valid if 
$\epsilon_F\gg \hbar \omega_T$, where $\omega_T$ is the frequency of the 
oscillator (which for the time being we assume to be isotropic). 
This condition is always satisfied if the 
particle number is sufficiently large, since for a harmonic potential
$\epsilon_F=(6N_{\sigma})^{1/3}\hbar \omega_T$, with $N_{\sigma}$ being the
number of particles of one species. 

Another condition for applicability of Eq. (\ref{tbcs})
 is that $k_BT_c\gg \hbar \omega_T$~\cite{georg}.
When this latter condition is not satisfied, 
the shell structure of the harmonic oscillator
is crucial when determining the superfluid properties of the gas, 
and Eq. (\ref{tbcs}) in general breaks down.

 In a superfluid Fermi gas at zero temperature the size of 
the vortex core of a singly quantized vortex
is given approximately by the BCS coherence length 
$\xi_{BCS}=\hbar v_F/\pi \Delta_0$, where $v_F=\hbar k_F/m_a$ 
is the Fermi velocity and $m_a$ 
the mass of a single atom. It is clear that in order for a vortex to
appear at all, the BCS coherence length (size of the vortex core)
at the center of the cloud has to be smaller
than the size of the cloud itself, which in the Thomas-Fermi approximation
is given by $R_{TF}=(2\epsilon_F/m_a\omega_T^2)^{1/2}$. If this were
not so the superfluid properties of the system would be more like those
of a nucleus (for which $\xi\gtrsim R$) than those of a bulk superfluid.

\begin{figure}[htbp]
  \begin{center}
    \psfig{file=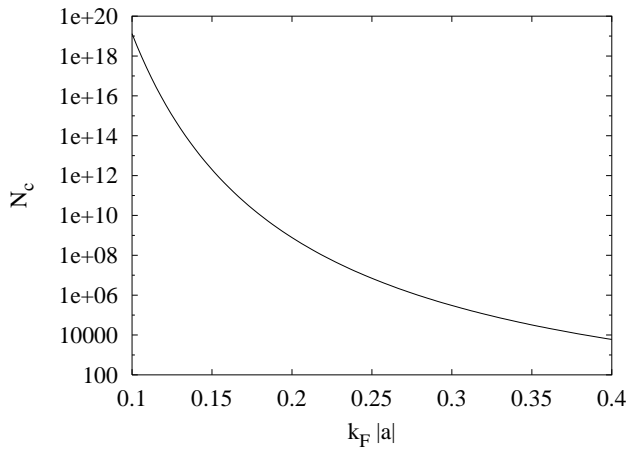,width=0.34\textwidth,angle=-90}
    \vspace{.2cm}
    \caption{Critical number of atoms per spin species 
for which $\xi_{BCS}/R_{TF}=1$ in an isotropic trap.
Well above the line the local density approximation, and thus 
Eq. (\ref{tbcs}), applies, below the line the system is intrinsically 
finite sized.}
    \label{fig:cohlength}
  \end{center}
\end{figure}

Substituting the appropriate expressions one can immediately see that 
$\xi_{BCS}/R_{TF}=\pi^{-1}\hbar\omega_T/\Delta_0$. So that requiring
$\xi_{BCS}\ll R_{TF}$ 
corresponds to demanding that  $\Delta_0\gg \pi^{-1}\hbar \omega_T$.
This condition is automatically satisfied if $k_BT_c\gg \hbar\omega_T$, but 
is not at all obviously realized in possible practical circumstances.
Indeed, if we assume the validity of equation (\ref{tbcs}), 
and of the related value of $\Delta_0$, and we use the expression 
$\epsilon_F=(6N_{\sigma})^{1/3}\hbar \omega_T$ for the Fermi energy, we 
obtain
\begin{equation}
N_{\sigma}\gg \frac{(e/2)^7}{6\pi^3}\;e^{3/\lambda}
\label{xioR}
\end{equation} 
We may then immediately see that unless $\lambda$, and therefore 
$k_F|a|$, is sufficiently close to one,
the exponential is very large and the condition in Eq. 
(\ref{xioR}) is not satisfied, implying that the coherence length 
is much larger than the radius of the cloud and the
rotation pattern very different from a vortex state.
If however $k_F|a|$ is too close to one ($\sim 0.3-0.4$ or more) the formula 
becomes unreliable because the
gas is no more dilute and effects due to induced interactions, 
which strongly modify the value of $\Delta_0$ obtained in 
the dilute limit, must be taken into account \cite{combescot}. For the
sake of the present work we shall not consider these effects; we  study 
 regions of densities in which Eq. (\ref{tbcs}) is reasonably 
reliable keeping  $k_F|a|\lesssim 0.4$. 
There is then a region of applicability of Eq. (\ref{tbcs}) for a trapped
gas which depends on the number of particles 
$N_{\sigma}$ and the scattering length.
In order to find this region we impose the equality in 
Eq. (\ref{xioR}), and we plot in figure \ref{fig:cohlength}
the critical number of atoms $N_{\sigma,c}$ for  which 
$\xi_{BCS}/R_{TF}=1$, as a function of $k_F|a|$.
Well above the curve we are in the regime where the local density 
approximation can be applied and a vortex may form,
and below it the superfluid has a character more
related to that of a nucleus.
Since the value of $k_F|a|$ can be simply increased by keeping 
the number of particles fixed and tightening
the external trapping potential,
we see that if $N_{\sigma}$ is sufficiently large 
($\gtrsim 10^5$) these systems have the
interesting possibility of going from one regime to the other.

In the remainder of this work we assume that we are in the upper region
of fig. \ref{fig:cohlength} and therefore that $\xi_{BCS}\ll R_{TF}$.
In this region a vortex forms in the cloud if it is stirred 
at an angular velocity greater than a critical one $\omega_{c1}$, which
we shall calculate using a thermodynamic approach.

\section{Vortex in a uniform gas}\label{Uniformvortex}

Let us for the time being suppose that the vortex we want to describe is
in a uniform gas. In particular we may take the system to be in a cylinder
of radius $R_c\gg \xi_{BCS}$.

Associated with the vortex there is a superfluid 
velocity flow which decreases with the distance from the vortex axis:
${\bf v}_v(\rho)={\bf e}_{\phi}\,\kappa\hbar/2m_a\rho$, where
$\kappa$ is the number of quanta of circulation of the vortex. 
In a simple model
this velocity field extends from $\rho \sim \kappa\xi_{BCS}$ to $\rho=R_c$.
At distances shorter than $\sim \kappa\xi_{BCS}$,
the kinetic energy associated with the rotation becomes
high enough to break the Cooper-pairs, and thus the fluid inside a cylinder
of radius $\sim \kappa\xi_{BCS}$ about the vortex axis can be thought of as being
in a normal (non-superfluid) state.
The energy per unit length associated with a vortex is then given 
by the sum of two 
contributions. One is the kinetic energy due to the flow: 
\begin{eqnarray}
\nonumber
{\cal{E}}_{kin}&=&
\int_{\kappa\xi_{BCS}}^{R_c} 2\pi\rho\; d\rho\;\; m_a n_{\sigma}
\left[\frac{\kappa\hbar}{2m_a\rho}\right]^2\\
&=&\frac{\pi\kappa^2\hbar^2n_{\sigma}}{2m_a}\ln\frac{R_c}{\kappa\xi_{BCS}}
\end{eqnarray}
and the other one is the loss in condensation energy about the vortex 
axis
\begin{eqnarray}
\nonumber
{\cal{E}}_{cond}&\sim& \pi\kappa^2\xi_{BCS}^2 \epsilon_{cond}\\
&=& \frac{\pi\kappa^2\hbar^2n_{\sigma}}{2m_a}\frac{3}{\pi^2}
\end{eqnarray}
where $\epsilon_{cond}=3\Delta_0^2 n_{\sigma}/4\epsilon_{F}$ 
is the condensation
energy per unit volume due to the pairing \cite{degennes}, and the usual
expression for $\xi_{BCS}$ has been employed.
Notice that we have here introduced the one-species particle density 
$n_{\sigma}$, and we have supposed that this is a
constant throughout the system, since contrary to the boson case
it is not the particle density but only the pairing field that changes close 
to the vortex axis~\cite{degennes}.

The total energy per unit length of a vortex is therefore given in the 
simple model by
\begin{equation}\label{vorten}
{\cal{E}}_{v}={\cal{E}}_{kin}+{\cal{E}}_{cond}
\simeq \frac{\pi\kappa^2\hbar^2n_{\sigma}}{2m_a} 
\ln\left(1.36\frac{R_c}{\kappa\xi_{BCS}}\right).
\end{equation}
The important feature of this result is to point out that for large
systems (i.e. for which $R_c\gg \kappa\xi_{BCS}$) the most relevant 
contribution is the logarithmic
one arising from the kinetic integration. The value of the constant inside 
the logarithm will  depend on the choice of the model used
to describe the vortex. A more reliable value would be 
obtained from a numerical solution of
the Bogoliubov-de Gennes equations, although it is unlikely that it
will differ significantly from the one found here, since 
one expects it in any case to be of order one. 
As an example of what a different approach may yield, 
in App.\ A we show the result for the total energy of the vortex
obtained using a zero-temperature Ginzburg-Landau model.
As we shall see one obtains, as foreseen, the same expression
as in Eq. (\ref{vorten}), with coefficient 1.65 instead of 1.36 inside
the logarithm.

For what follows we shall not need to know the precise value of this
coefficient, which may be better determined in the future, and we shall 
therefore leave it unspecified and state our result as 
\begin{equation}\label{vortenII}
{\cal{E}}_{v}\simeq \frac{\pi\kappa^2\hbar^2n_{\sigma}}{2m_a} 
\ln\left(D\frac{R_c}{\kappa\xi_{BCS}}\right)
\label{Fv}
\end{equation}
with the understanding that $D$ is some contant of order one.
From Eq. (\ref{vortenII}) it is already clear that one
vortex with $\kappa=\tilde{\kappa}\neq 1$ has greater energy than 
$\tilde{\kappa}$ vortices with $\kappa=1$ since in any case we need to 
have $\kappa\xi_{BCS}\ll R_c$. This implies that vortices with
$\kappa\neq 1$ are unstable \cite{ginzpit}. 
Therefore for the considerations that follow
we shall set $\kappa=1$.

With this solution, recalling that the thermodynamic 
critical velocity for formation of a first vortex is given by
$\omega_{c1}={\cal{E}}_{v}/{\cal L}_{v}$ \cite{nozpinesII}, 
and using the fact that the total angular
momentum per unit length of the system with a vortex is 
${\cal L}_v\simeq \hbar \pi R_c^2 n_{\sigma}$,
corresponding to $\hbar$ per Cooper pair,
we can immediately state what the 
critical velocity is in a uniform system, which
is of course a well known result
\begin{equation}
\omega_{c1}=\frac{\hbar}{2m_a R_c^2}\ln\left(D\frac{R_c}{\xi_{BCS}}\right).
\end{equation}
The result should be compared with the critical velocity found 
for a Bose-Einstein condensate. This is completely analogous if
the mass of a single bosonic atom is replaced with that of a Cooper pair
($2m_a$), and the boson coherence length by the BCS 
one.

Notice that since $\xi_{BCS}\propto \Delta_0^{-1}$, from the measurement
of $\omega_{c1}$ one could in principle deduce the value of $\Delta_0$
if $D$ is known.
This possibility is usually lost however in a non-uniform system since 
several values of $\Delta_0$ are integrated over.

\section{Vortex formation in trapped gases}\label{Trappedvortex}

In this section we calculate the energy of the vortex in a trapped gas,
specializing to the trapping configurations used in a typical experiment.

The atoms are generally confined in a cylindrically symmetric 
harmonic potential of the form 
\begin{equation}\label{trap}
V_{ext}({\bf r})=\frac{1}{2}m\omega_z^2[z^2+\lambda_T^2(x^2+y^2)]
\end{equation}
and the density profile of the gas is, within the Thomas-Fermi 
approximation, given by  
\begin{equation}
n_{\sigma}(\rho,z)=n_{\sigma,0}\left(1
-\frac{\lambda_T^2\rho^2+z^2}{R_z^2}\right)^{3/2}.
\label{nTF}
\end{equation}
Here $n_{\sigma,0}=n_{\sigma}(0,0)$ is the density at the center of the cloud
and we have taken the 
profiles of the two species to be identical. The anisotropy of the trap  is
 controlled by the coefficient $\lambda_T$.
The energy of a cloud with a vortex along the $z$ axis can be 
calculated with the procedure devised by Lundh {\em et al.} \cite{lundh}.
One can divide the cloud in vertical slices
of height $dz$ and use the result (\ref{Fv}) for a cylinder of
radius $\rho_1$ such that $\xi_{BCS}\ll \rho_1\ll R_{\perp}=R_z/\lambda_T$, 
within which one can assume that the gas is approximately uniform.
The energy per unit length associated with the vortex in a slice at $z$ 
is then given by
\begin{eqnarray}\label{Venergy}
\nonumber
{\cal E}_{v}(z)=\frac{\pi\hbar^2n_{\sigma}(0,z)}{2m_a}
\ln\left(D\frac{\rho_1}{\xi_{BCS}(z)}\right)\nonumber\\
+\int_{\rho_1}^{R_{\perp}(z)}2\pi\rho\; d\rho\;
m_a n_{\sigma}(\rho,z)\left[\frac{\hbar}{2m_a\rho}\right]^2,
\end{eqnarray}
where $R_{\perp}(z)=(1-z^2/R_z^2)^{1/2}R_z/\lambda_T$ is the value of 
$\rho$ up to which the cloud extends for a given
$z$, and  $n_{\sigma}(0,z)$ is the density on the $z$-axis at height $z$. 
The  second term in Eq.\ (\ref{Venergy}) gives the kinetic energy 
of the superfluid outside the cylinder of radius $\rho_1$.

With $n_{\sigma}(\rho,z)$ given by (\ref{nTF}) we then get
\begin{eqnarray}
\nonumber
{\cal E}_{v}(z)=\frac{\pi\hbar^2n_{\sigma,0}}{2m_a}\left[
\frac{n_{\sigma}(0,z)}{n_{\sigma,0}}
\ln\left(D\frac{\rho_1}{\xi_{BCS}(z)}\right)\right.\nonumber\\
\left.+\int_{\rho_1}^{R_{\perp}(z)} \left(1
-\frac{\lambda_T^2\rho^2+z^2}{R_z^2}\right)^{3/2}\frac{d\rho}{\rho}\right].
\end{eqnarray}
This result differs from the boson case in ref. \cite{lundh} 
in the power 3/2 instead of 1 
in the density distribution. Using the fact that
\begin{eqnarray}
\nonumber
\int (1-x^2)^{3/2}\frac{dx}{x}&=&\sqrt{1-x^2}
+\ln\left(\frac{x}{1+\sqrt{1-x^2}}\right)
\\
&+&\frac{1}{3}(1-x^2)^{3/2}
\end{eqnarray}
and that unless $z$ is very close to $R_z$ one can assume 
$\rho_1\ll R_{\perp}(z)$, we finally obtain
\begin{equation}
{\cal E}_{v}(z)=\frac{\pi\hbar^2n_{\sigma,0}}{2m_a}\left(1-\frac{z^2}{R_z^2}
\right)^{3/2}\ln\left(\frac{2}{e^{4/3}}D\frac{R_{\perp}(z)}{\xi_{BCS}(z)}\right).
\label{Evz}
\end{equation}
In order to proceed with the $z$ integration we need to know the explicit
dependence of $\xi_{BCS}$ on $z$.
In the dilute gas approximation where Eq.\ (\ref{tbcs}) is valid
this is given by 
\begin{eqnarray}
\nonumber
\xi_{BCS}(z)= 
\frac{2}{\pi}\left(\frac{e}{2}\right)^{7/3}
\left(1-\frac{z^2}{R_z^2}\right)^{-1/2}\nonumber\\
\times\exp\left[\frac{1}{\lambda_0}\left(1-\frac{z^2}{R_z^2}\right)^{-1/2} 
\right]k_{F,0}^{-1},
\end{eqnarray}
where $k_{F,0}=(2m_a \epsilon_F/\hbar^2)^{1/2}$ and 
$\lambda_{0}=2k_{F,0}|a|/\pi$ are the local 
Fermi wave-number and $\lambda$ respectively,
evaluated at the center of the cloud.
Inserting this value into Eq. (\ref{Evz}), using the expression for 
$R_{\perp}(z)$ and integrating over $z$ we get after some cumbersome but
straightforward calculations 
\begin{equation}\label{energy}
E_{v}=\frac{\pi\hbar^2n_{\sigma,0}}{2m_a}\frac{4}{3}R_z
\left[\frac{9\pi}{32}\ln\left(\frac{2^{4/3}\pi D}{e^{5/2}}\cdot
\frac{\epsilon_F}{\hbar\omega_{\perp}}\right)
-\frac{1}{\lambda_{0}}\right]
\end{equation}
Note that Eq.\ (\ref{energy}) 
predicts the energy cost of the vortex to be negative for 
small $k_{F,0}|a|$ and $\epsilon_F/\hbar\omega_{\perp}$ not too large.
 This is clearly an unphysical result reflecting the fact 
that in the limit of relatively few particles trapped and
small  $k_{F,0}|a|$, the condition $\xi_{BCS}\ll R_{\perp}$ 
is violated making Eq.\ (\ref{energy})
invalid. In the regime  $\xi_{BCS}\ll R_{\perp}$, Eq.\ (\ref{energy}) 
yields positive vortex energies as expected. 
If we ignore the non-rotating particles at the core of the vortex, 
the total angular momentum of the vortex state is $L_v=N_\sigma\hbar$ 
and the critical rotation frequency $\omega_{c1}=E_v/L_v$
for the formation 
of a vortex in a trap given by Eq.\ (\ref{trap}) is
\begin{equation}\label{criticalvortex}
\omega_{c1}=\omega_{\perp}\frac{16}{3\pi}
\frac{l^2_{\perp}}{R_{\perp}^2}
\left[\frac{9\pi}{32}\ln\left(\frac{2^{1/3}\pi D}{e^{5/2}}\cdot
\frac{R_{\perp}^2}{l_{\perp}^2}\right)
-\frac{1}{\lambda_{0}}  \right]
\end{equation}
with $l_{\perp}=\sqrt{\hbar/m\omega_{\perp}}$ being the harmonic 
oscillator length in the radial direction. 
For realistic parameters of the gas, this critical frequency 
is rather small: choosing for $D$ the value obtained in App.\ 
\ref{Ginzburg}, taking $k_{F,0}|a|=0.4$, 
$\omega_{\perp}=\omega_z=\omega_T$ (i.e. $\lambda_T=1$) and 
$\epsilon_F=200\,\hbar\omega_T$ corresponding to 
isotropic trap with $N_\sigma\sim 1.3\times10^6$, we obtain
 $\omega_{c1}\simeq0.0035\,\omega_{\perp}$. 
The reason for the critical frequency being so 
small is that the angular momentum per atom is $\hbar/2$ 
yielding $L_v=N_\sigma\hbar$
whereas the energy given by Eq.\ (\ref{energy}) 
only scales as  $N_\sigma^{2/3}$. 
The Fermi pressure expands the cloud and reduces the 
density at the center of the 
trap. Since the energy of the vortex mainly comes from 
regions close to the vortex axis 
where the superfluid velocity  is high, the energy of 
the vortex is correspondingly 
reduced.

\section{Observation of the vortex}\label{Observation}

Contrary to the situation for 
Bose-Einstein condensates, the presence of a vortex in the Fermi
 gas does not alter the density profile significantly~\cite{degennes}. 
One cannot therefore observe the vortex simply by looking 
at the density profile. It has been suggested to use 
the laser probing method of Ref.\ \cite{Laser} to detect
the local decrease of the pairing near the center of 
the vortex~\cite{Rodriguez}. Here we examine
a different method based on measuring the collective 
mode spectrum of the gas. In the case of no 
vortex present,  excitations of the gas  carrying equal and
opposite angular momentum along the $z$-axis are degenerate in energy. 
The velocity field associated with a vortex aligned with this axis
lifts the degeneracy since the rotational symmetry is removed;
the velocity flow of the excitation is either parallel 
or anti-parallel to that of 
the vortex giving rise to an energy splitting of the 
modes~\cite{Stringari,Sinha,Svidzinsky}. Since the 
collective mode frequencies  of the gas can be measured 
with a fairly high precision, the possibility of detecting
 the presence of the vortex by its spectroscopic signatures 
is a promising method.
 Indeed, this method has proven to be very useful in the case 
of a vortex in a BEC~\cite{Chevy}. 
The calculations will be carried out for an isotropic trap with
$R_{\perp}=R_z=R_{TF}$ and $\omega_{\perp}=\omega_z=\omega_T$.

In the $\xi_{BCS}\ll R_{TF}$ limit considered in this paper, 
the collective modes of the superfluid 
gas for $T=0$ can be calculated using a hydrodynamic theory. 
The relevant continuity and 
 superfluid velocity equations read~\cite{Landau}:
\begin{eqnarray}\label{2fluid}
\partial_t n({\mathbf{r}},t)&=&-\nabla{\mathbf{j}}({\mathbf{r}},t)
\nonumber\\
\partial_t {\mathbf{v}}_s({\mathbf{r}},t)&=&-\frac{1}{m_a}
\nabla[m_a{\mathbf{v}}_s^2/2+\mu_F+V_{ext}]
\end{eqnarray}
with $n_s({\mathbf{r}},t)$, $n_n({\mathbf{r}},t)$, and 
$n({\mathbf{r}},t)=n_s({\mathbf{r}},t)+n_n({\mathbf{r}},t)$
 being the superfluid-, normal- and   total density of the 
gas respectively. The total current is 
 ${\mathbf{j}}({\mathbf{r}},t)=n_s({\mathbf{r}},t)
{\mathbf{v}}_s({\mathbf{r}},t)+n_n({\mathbf{r}},t)
{\mathbf{v}}_n({\mathbf{r}},t)$,
 ${\mathbf{v}}_s({\mathbf{r}},t)$ is the superfluid velocity 
and ${\mathbf{v}}_n({\mathbf{r}},t)$ the velocity of the 
normal fluid. For  $\xi_{BCS}\ll R_{TF}$, the extent of the
 vortex core is small compared to the size of the cloud, and 
the main effect of the vortex on the collective mode spectrum 
is the presence of the vortex velocity field 
${\mathbf{v}}_v({\mathbf{r}})={\mathbf{e}}_\phi\,\kappa\hbar/2m_a\rho$.
In these considerations we keep the winding number $\kappa$ explicitly
different from one for generality. 
Writing $n({\mathbf{r}},t)=n_0({\mathbf{r}})+\delta n({\mathbf{r}},t)$ and 
 $ {\mathbf{v}}_s({\mathbf{r}},t)= {\mathbf{v}}_v({\mathbf{r}})+ 
{\mathbf{u}}({\mathbf{r}},t)$, where $n_0({\mathbf{r}})$ is the 
equilibrium density profile with the vortex alone (which we take to be
coincident with the Thomas-Fermi one without vortex), and
linearizing  in $\delta n({\mathbf{r}},t)$ and 
 ${\mathbf{u}}({\mathbf{r}},t)$, Eq.\ (\ref{2fluid}) can be written as 
\begin{eqnarray}
\nonumber
\left(\omega-\frac{\kappa\hbar m}{2m_a\rho^2}\right)^2 &&
m_a n_0({\mathbf{r}})
\left[\frac{\partial n_0({\mathbf{r}})}
{\partial P_0({\mathbf{r}})}\right]_{T=0}\Phi({\mathbf{r}},t)\\
&=&-\nabla\cdot[n_0({\mathbf{r}})\nabla\Phi({\mathbf{r}},t)].
\label{Hydrofinal}
\end{eqnarray}
Here, $\omega$ is the frequency of the collective mode, 
$P_0({\mathbf{r}})$ is the equilibrium pressure profile, and
$m$ is the magnetic quantum number of the mode.
The velocity field associated with the mode has been written as 
${\mathbf{u}}({\mathbf{r}},t)=\nabla\Phi({\mathbf{r}},t)$ 
with $\Phi({\mathbf{r}},t)=\Phi(r,\theta)\exp[i(m\phi-\omega t)]$. 
The term $\kappa\hbar m/2m_a\rho^2$ in 
Eq.\ (\ref{Hydrofinal}) comes from the presence of the vortex 
velocity field ${\mathbf{v}}_v$. Without this term, 
Eq.\ (\ref{Hydrofinal}) has been solved for a spherical 
symmetric trap by writing  
 $\Phi_{nlm}({\mathbf{r}})=\Phi_{nl}(r)Y_{lm}(\theta,\phi)$ 
yielding the spectrum 
${\omega_{nl}}_0=2\omega_T\sqrt{(n^2+2n+ln+3l/4)/3}$ with 
$n=0,1,2,\ldots$~\cite{sphericalhydro}.
  From Eq.\ (\ref{Hydrofinal}), we see that the frequency 
shift of a given mode induced by the vortex can be 
calculated perturbatively as 
\begin{equation}\label{perturbative}
\omega^2_{nlm}-\omega^2_{nl0}=\frac{\kappa\hbar m\omega_{nl0}}{m_a}
\frac{\langle\Phi_{nlm}|\rho^{-2}|\Phi_{nlm}
\rangle}{\langle\Phi_{nlm}|\Phi_{nlm}\rangle}.
\end{equation}
Here $\langle\Phi_{nlm}|f({\mathbf{r}})|\Phi_{nlm}\rangle$ 
denotes the spatial average  
$\int _0^{R_z}w(r)dr\int d\Omega \Phi^2_{nlm}({\mathbf{r}})f({\mathbf{r}})$ 
with the weight 
function $w(r)=r^2(1-r^2/R^2_{TF})^{1/2}$. This anomalous
weight has to be introduced in place of simply $r^2$
because the operator in Eq. (\ref{Hydrofinal}) without the perturbation
is not Hermitian.

As pointed out in Ref.\ \cite{Svidzinsky}, the perturbative 
procedure works for $|m|\ge2$; for $|m|<2$ it predicts 
an unphysical $\rho\rightarrow 0$ divergence in the
 density fluctuation of the mode.   With no vortex present, the lowest 
mode for a given angular momentum is the surface mode 
$\Phi_{n=0lm}(r)\propto r^l$ with  frequency
 $\sqrt{l}\omega_T$.
Recalling that $\rho=r\sin\theta$ and using the fact that 
$(4\pi)^{-1}\int d\Omega |Y_{lm}(\Omega)|^2/\sin^2\theta=(2l+1)/2|m|$, 
the matrix elements in Eq.\ (\ref{perturbative}) 
can be calculated analytically for these surface modes and we obtain 
for the frequency shift:
\begin{equation}\label{splitting}
\frac{\omega_{lm}^2-\omega^2_{l0}}{\omega^2_{l0}}=
\text{sgn}(m)\frac{\kappa(l+2)}{2\sqrt{l}(6N_\sigma)^{1/3}}
\end{equation}
with $|m|\ge2$.
As expected, the vortex splits the $2l+1$ degenerate modes 
depending on the direction of the projection of their 
angular momentum on the $z$-axis. 
Not all the modes are split however since the splitting is 
independent of $|m|$ in analogy with the equivalent 
result for bosons~\cite{Svidzinsky}. 
Particularly important is the result 
$(\omega_{2,\pm 2}^2-\omega^2_{20})/\omega^2_{20}=
\pm \sqrt{2}\kappa/(6N_{\sigma})^{1/3}$
for the quadrupolar mode 
$l=2$, $m=\pm 2$, since this mode is easily excited in trapped gases
and has been already employed for a precise determination of the critical
frequency for vortex nucleation in Bose gases.

The same result can be obtained following the sum rule approach of ref.
\cite{Stringari}. From that the splitting is found to be given by
$\omega_{2,2}-\omega_{2,-2}=2<l_z>/(m_a<\rho^2>)$, with $<l_z>$
expectation value of the angular momentum along the $z$-axis per atom
($\kappa\hbar/2$ in the case of a vortex), 
and $<\rho^2>$ expectation value of $x^2+y^2$ (equal to $R_{TF}/4$ for an
isotropic cloud with a Thomas-Fermi density profile).
From the latter result one can immediately see that 
the splitting of the modes of a Fermi superfluid
is in general smaller compared with the BEC case, the reason being that
given the same number of atoms the radius 
of a fermionic cloud is usually larger due to the Pauli repulsion and thus
the expectation value of $r_{\perp}^2$ is also correspondingly larger,
and the splitting reduced.  
For $2N_\sigma=10^6$ particles trapped, the $m=\pm2$ 
quadrupole modes are split by $\sim 1\%$. 
Although this is a rather small shift, it should be 
measurable assuming the same high spectroscopic 
precision demonstrated for BEC's can be obtained for 
trapped Fermi gases~\cite{Jin}.

\section{Conclusion}\label{Conclusion}

In this paper we considered various aspects of the 
vortex state of a dilute superfluid Fermi gas at $T=0$. 
For a trapped system, we found that a large number of particles 
and a not too small scattering length yields $\xi_{BCS}\ll R_{TF}$ and 
the vortex is well confined within the gas. 
We then used a simple model to calculate the energy of a 
vortex in a uniform medium.
Subsequently, using the fact that the structure of the 
vortex near the rotation axis is essentially unaffected by the trapping 
potential we derived an expression for the vortex energy 
in a trap, and we employed this energy expression to calculate 
the thermodynamic critical rotation frequency for the formation 
of a vortex. 
Finally, we suggested a way of observing the presence 
of the vortex by calculating 
perturbatively its influence on the collective mode spectrum of the gas.
In the Appendix we report an alternative, less na\"\i ve, description 
of the vortex in a uniform medium and find a slightly different value for
its energy compared the one obtained in Sec. \ref{Uniformvortex}.

\section{Acknowledgment}
We would like to acknowledge valuable discussions with C.\ J.\ Pethick.

\appendix

\section{A Ginzburg--Landau description of the  vortex core}

\label{Ginzburg}
We here present a  Ginzburg--Landau 
description of the vortex core in a uniform gas, and the consequent result
for the total energy of a vortex in a cylindrical bucket of radius $R_c$.
As is well-known, Ginzburg--Landau
theory is only valid for temperatures such that $|T-T_c|/T_c\ll1$ 
but  the following calculation can   
be used  for a qualitative estimate at 
$T=0$~\cite{epstbaym}.

The extension of the Ginzburg-Landau theory to zero temperature for a uniform
system can be done by imposing to the free energy 
\cite{fetter}
\begin{eqnarray}
\nonumber
F_{GL}&=&\int d^3r f_{GL}({\mathbf{r}})\\
&=&\int \left[\frac{\hbar^2|\nabla\psi|^2}{4m_a}+A|\psi|^2
+\frac{B}{2}|\psi|^4\right]d^3r,
\label{FGL}
\end{eqnarray}
to be equal to the condensation energy density, which in a uniform system
is given by $\epsilon_{cond}=-3\Delta_0^2n_{\sigma}/4\epsilon_F$ 
\cite{degennes}.
$\psi$ is here the well known Ginzburg-Landau order parameter.
Upon minimization of (\ref{FGL}) with respect to $\psi^*$ we obtain 
the Ginzburg-Landau equation
\begin{equation}
-\frac{\hbar^2}{4m_a}\nabla^2\psi+A\psi+B|\psi|^2\psi=0.
\label{EGL}
\end{equation}
For a uniform system the solution is $|\psi_0|^2=-A/B$ and from the fact 
that the Ginzburg--Landau free energy then coincides with the condensation one 
$f_{GL}=-3\Delta_0^2n_{\sigma}/4\epsilon_F$,
we obtain $A=-3\Delta_0^2/2\epsilon_F$ and $B=-A/n_{\sigma}$.

We now calculate  the structure and the energy of the vortex.
A vortex along the $z$-axis is described by writing the order 
parameter in cylindrical coordinates as $\psi({\bf r})=f(\rho)
e^{i\kappa\phi}$.
Replacing this expression into Eq. (\ref{EGL}) we find
\begin{equation}
-\frac{1}{x}\frac{d}{d x}\left(x\frac{d\chi}{d x}\right)
+\frac{\kappa^2}{x^2}\chi+\chi^3-\chi=0,
\end{equation}
where we introduced the 
dimensionless quantities $\chi=f/|\psi_0|$ and $x=\rho/\xi_{GL}$. 
We used the fact that $f$ does not vary along the $z$ direction if the 
system is uniform, and we defined the Ginzburg--Landau coherence 
length $\xi_{GL}^2=\hbar^2/4m_a A$, which 
implies $\xi_{GL}=\hbar v_F/2\sqrt{3}\Delta_0=0.907 \xi_{BCS}$.
This equation has exactly the same form as the Gross-Pitaevskii 
equation for a 
vortex in a uniform boson cloud~\cite{ginzpit,pethicksmith}. 
It can be solved numerically and the results for the lowest $\kappa$
($\kappa=1,2,3$) was first obtained
by Ginzburg and Pitaevskii \cite{ginzpit}. A very good approximate 
solution for $\kappa=1$ can be obtained by a variational calculation yielding 
$\chi=x/(2+x^2)^{1/2}$~\cite{Fettervar}.
Using this solution in Eq.\ (\ref{FGL}), 
one finds that the energy cost associated with the  vortex core 
 is given by
\begin{equation}
{\cal E}_{v}(z)=\pi\frac{\hbar^2}{2m_a}n_{\sigma}
\ln\left(\frac{e^{3/4}}{\sqrt{2}}\frac{R_c}{\xi_{GL}}\right).
\end{equation}
This result is now identical 
with that for a Bose-Einstein condensate, with
the mass of a single bosonic atom replaced by that of a Cooper pair ($2m_a$),
and the boson coherence length replaced by the 
Ginzburg--Landau one. Using $\xi_{GL}=0.907 \xi_{BCS}$, we
obtain  ${\cal E}_{v}=\pi\hbar^2n_{\sigma}\ln(D\, R_c/\xi_{BCS})/2m_a$  
with $D=1.65$.


\end{document}